\documentclass[]{IAC_style}
\usepackage{amsmath}
\usepackage{amsfonts}
\usepackage{bm}
\usepackage{commath}

\usepackage{algorithmic}
\usepackage{algorithm}
\usepackage{array}
\usepackage{caption}
\usepackage{subcaption}
\usepackage{textcomp}
\usepackage{stfloats}
\usepackage{url}
\usepackage{verbatim}
\usepackage{graphicx}
\usepackage{cite}
\usepackage{comment}
\usepackage[super]{nth}
\usepackage[version=4]{mhchem}
\usepackage[utf8]{inputenc}
\usepackage{siunitx}
\usepackage{longtable,tabularx}
\usepackage{hyperref}
\usepackage{tikz}
\usetikzlibrary{arrows.meta,calc}

\DeclareMathOperator{\Exp}{Exp}
\DeclareMathOperator{\Log}{Log}

\begin{document}

% -------------------------------------------------------------------------
% IAC 2026 front matter
% -------------------------------------------------------------------------

\title{Robust LEO Attitude Determination via Sun–Horizon Fusion Using Factor Graph Optimization}

% Author list and affiliations
\IACauthor{Sabire Betül Çınar}{1}{0}
\IACauthor{Abdülbaki Şanlan}{1}{0}
\IACauthor{Ege Ç. Altunkaya$^{\orcidlink{0000-0001-8938-7632}}$}{2}{0}
\IACauthor{Emre Koyuncu$^{\orcidlink{0000-0002-0726-4979}}$}{1}{0}
\IACauthor{İbrahim Özkol$^{\orcidlink{0000-0002-9300-9092}}$}{1}{0}

\IACauthoraffiliation{
Faculty of Aeronautics and Astronautics,
Istanbul Technical University,
Istanbul, Türkiye, 34469.
}

\IACauthoraffiliation{
Aviation Institute,
Istanbul Technical University,
Istanbul, Türkiye, 34469.
}

\abstract{
Accurate attitude determination is essential for small satellites in Low Earth Orbit (LEO). A low-cost is to fuse gyroscope and Sun sensor measurements with the nadir direction measured from the Earth horizon by a visible-spectrum camera. However, the horizon in RGB images is degraded by atmospheric scattering, clouds, sun glare, and motion blur, so that the accuracy of the nadir measurement varies from image to image. Conventional image-based horizon sensors extract the horizon with fixed thresholds and edge detectors and assume a constant measurement noise, which assigns the same confidence to degraded and accurate measurements. This work presents an attitude determination framework in which a neural network not only detects the horizon but also predicts how accurately it is localized. A U-Net segments the Earth in the camera image, and an uncertainty head trained with a Gaussian negative log-likelihood loss predicts the standard deviation of the horizon location at each pixel. These per-pixel uncertainties weight the least-squares fit of the horizon cone and are propagated into the covariance of the measured nadir direction, so that each image provides a nadir measurement together with its own covariance and the attitude estimation becomes robust to degraded images. The horizon measurements are fused with the Sun sensor and gyroscope measurements in a sliding-window factor graph optimization, which estimates the attitude on SO(3) together with the gyroscope bias. Rather than using an conventional extended Kalman filter, Sliding Window Factor Graph Optimization  
is used to naturally incorporates sensors operating at different with varying updated measurement covariances.
% TODO: add one sentence summarizing the quantitative simulation results.
}

\maketitle
\thispagestyle{fancy}


\section{Introduction}
\label{introduction}

Attitude determination is a fundamental function of a spacecraft, since the pointing of its payloads, antennas, and solar arrays relies on the knowledge of its orientation~\cite{MarkleyCrassidis2014}. Small satellites in Low Earth Orbit (LEO) typically determine their attitude by combining low-cost sensors, including Sun sensors~\cite{ortega2010miniaturized,janson2012attitude}, magnetometers, gyroscopes, and Earth horizon sensors~\cite{astheimer1971high}. Since a single direction measurement constrains only two of the three rotational degrees of freedom, the attitude is determined from at least two non-collinear directions.

Earth horizon sensors measure the nadir direction by detecting the boundary between the Earth and space. Since the early days of spaceflight, they have mostly operated in the long-wave infrared~\cite{astheimer1971high,christian2021tutorial}, where the Earth is bright during both sunlight and eclipse~\cite{nguyen2018attitude}, and the carbon dioxide absorption band of the atmosphere yields a stable horizon profile~\cite{astheimer1971high}. On small satellites, static infrared thermopiles have been flown together with Sun sensors~\cite{janson2012attitude} and used to compute the nadir direction analytically~\cite{nguyen2018attitude}. Imaging sensors, in contrast, observe many horizon points, from which the nadir direction can be estimated directly~\cite{christian2016noniterative,christian2021tutorial}, and Deng et al.~\cite{deng2023large} applied such an approach to a wide field-of-view infrared camera with the same localization noise assumed for all horizon points. However, dedicated infrared sensors increase the cost and development time of a mission.

Visible-spectrum cameras, on the other hand, are inexpensive and offer higher resolution~\cite{christian2021tutorial}. Saadat~\cite{saadat2022attitude} repurposed a spacecraft inspection camera as a horizon sensor, and Braun and Barf~\cite{braun2023image}, whose work is the closest to ours, fused camera-based nadir and Sun directions with inertial and GNSS measurements in an error-state Kalman filter. The main difficulty of visible-spectrum horizon sensing is that the atmosphere does not produce a sharp brightness discontinuity at the horizon, but scatters sunlight and creates an apparent unpredictable horizon thickness~\cite{christian2021tutorial}. In addition, clouds, sun glare, and spacecraft structures in the field of view corrupt the detected horizon~\cite{braun2023image,saadat2022attitude}, so that the accuracy of the measured nadir direction varies considerably from image to image. However, both approaches extract the horizon with fixed thresholds, and neither measurement model reflects the quality of the individual image.

In parallel, deep learning has improved both the extraction of information from images and the quantification of its uncertainty. Bayesian deep learning enables a network to predict the uncertainty of its outputs~\cite{gal2016dropout,kendall2017uncertainties}, and such learned uncertainties have been shown to improve the robustness and accuracy of pose graph optimization in visual odometry~\cite{qiu2025mac}. For horizon detection onboard aircraft, deep semantic segmentation has recently shown higher robustness under clouds and glare~\cite{kapuscinski2026vision}, but it does not indicate when its prediction is unreliable.

For the fusion of attitude measurements, the extended Kalman filter (EKF) has long been the workhorse of spacecraft attitude estimation~\cite{lefferts1982kalman,crassidis2007survey}. Factor graph optimization (FGO), in contrast, jointly estimates the states within a time window and iteratively relinearizes the measurement models, which improves accuracy over the EKF and naturally accommodates asynchronous sensors~\cite{sibley2010sliding,wen2021factor,taylor2024factor}.

Table~\ref{tab:LiteratureComparison} summarizes the horizon-based approaches most closely related to this work. To the best of our knowledge, no previous work learns the horizon localization uncertainty from the image and propagates it into the measurement covariance of an attitude estimator.

To address this gap, this paper considers a small satellite in LEO equipped with a three-axis gyroscope, Sun sensors mounted on its faces, and a body-fixed RGB camera that observes the Earth horizon. The objective is to estimate the attitude and the gyroscope bias of the spacecraft from the asynchronous gyroscope, horizon, and Sun measurements while accounting for the image-dependent accuracy of the horizon measurements. The spacecraft position is assumed to be known, e.g., from a GNSS receiver, so that the reference nadir direction and, together with the solar ephemeris, the reference Sun direction are available in the Earth-fixed frame. Moreover, the Earth is modeled as a sphere, the camera is calibrated with a known mounting relative to the spacecraft body, and the spacecraft operates in sunlight with the illuminated horizon in the camera field of view, since neither the visible-spectrum camera nor the Sun sensors can operate in eclipse. Missions that require attitude determination during eclipse would need additional sensors, which are beyond the low-cost scope of this study.

\begin{table*}[!t]
    \centering
    \caption{Comparison of horizon-based approaches related to this work.}
    \label{tab:LiteratureComparison}
    \footnotesize
    \begin{tabularx}{\textwidth}{l l X X l}
        \hline
        Work & Spectrum & Horizon extraction & Horizon measurement uncertainty & Estimation \\
        \hline
        Nguyen et al.~\cite{nguyen2018attitude} & Infrared & Thermopile readings & Not modeled & Single epoch \\
        Deng et al.~\cite{deng2023large} & Infrared & Horizon points in image & Constant pixel noise & Single epoch \\
        Saadat~\cite{saadat2022attitude} & Visible & Fixed threshold, Canny edges & Not modeled & Single epoch \\
        Braun and Barf~\cite{braun2023image} & Visible & Fixed threshold, contour tracing & Constant pixel noise & Error-state Kalman filter \\
        Kapuscinski and Szczerba~\cite{kapuscinski2026vision} (aircraft) & Visible & Deep segmentation & Not modeled & Single epoch \\
        This work & Visible & Deep segmentation & Learned per-pixel uncertainty & Sliding-window FGO \\
        \hline
    \end{tabularx}
\end{table*}

% \begin{figure}[htb!]
%     \centering
%     \includegraphics[width=\columnwidth]{pics/example.jpg}

%     \caption{Overview of the considered problem. The spacecraft attitude is estimated from the gyroscope, the Sun sensors, and the nadir direction measured by an RGB horizon camera, whose accuracy varies with the imaging conditions.}
%     \label{fig:problemStatement}
% \end{figure}

The main contributions of this study are summarized as follows:

\begin{itemize}

    \item An uncertainty-aware horizon sensor for visible-spectrum cameras, in which a neural network detects the horizon and predicts its localization uncertainty at each pixel, and these uncertainties are propagated into the covariance of the measured nadir direction. Unlike the constant noise models in~\cite{braun2023image,deng2023large}, this covariance varies from image to image.

    \item A sliding-window FGO that fuses the uncertainty-aware horizon measurements with Sun sensor and gyroscope measurements to estimate the attitude and the gyroscope bias, as an alternative to the EKF commonly used for spacecraft attitude estimation.

    \item A publicly available dataset of real and synthetic RGB horizon images with manually labeled Earth regions.

\end{itemize}

The remainder of the paper is organized as follows. Section~\ref{preliminaries} introduces the notation, the attitude kinematics, and the fundamentals of FGO. Section~\ref{methodology} presents the horizon detection network, the uncertainty-aware nadir measurement, the Sun sensor model, and the sliding-window estimator. Section~\ref{Sec:Results} describe the simulation environment and present results, and Section~\ref{Sec:Conclusion} concludes the paper.

\section{Preliminaries}
\label{preliminaries}

In this section, we establish the notation used throughout the paper and present the attitude kinematics, the sensor measurement models, and the fundamentals of factor graph optimization.

\subsection{Notations}

A rotation matrix $\bm{R}_{\alpha}^{\beta} \in \mathrm{SO}(3)$ transforms vector coordinates from frame~$\mathcal{F}_\alpha$ to frame~$\mathcal{F}_\beta$, and $\bm{q}_{\alpha}^{\beta}$ denotes the corresponding unit quaternion, i.e., $\bm{R}_{\alpha}^{\beta} = \bm{R}(\bm{q}_{\alpha}^{\beta})$. A vector $\bm{v}_{\alpha\beta}^{\gamma}$ denotes a quantity of frame~$\mathcal{F}_\beta$ with respect to frame~$\mathcal{F}_\alpha$, resolved in frame~$\mathcal{F}_\gamma$. Direction vectors that are not associated with a pair of frames, such as the Sun and nadir directions, carry only the superscript of the resolving frame, e.g., $\bm{s}^{b}$. The frames used in this work are: $i$~(Earth-centered inertial frame aligned with the International Celestial Reference Frame, ICRF), $e$~(Earth-Centered Earth-Fixed, ECEF), $b$~(spacecraft body), and $c$~(camera, whose $x$- and $y$-axes point right and down in the image plane and whose $z$-axis lies along the optical axis).

Scalars are denoted by italic letters, vectors by bold lowercase letters, and matrices by bold uppercase letters. A quantity evaluated at the discrete time instant $t_k$ carries the subscript $k$, and $\dot{(\cdot)}$ denotes the time derivative. The accents $\bar{(\cdot)}$, $\hat{(\cdot)}$, and $\tilde{(\cdot)}$ denote nominal, estimated, and measured quantities, respectively, while $\delta(\cdot)$ denotes an error quantity. The $n \times n$ identity matrix is denoted by $\bm{I}_n$, and the $n \times m$ zero matrix by $\bm{0}_{n \times m}$, with $\bm{0}_n \triangleq \bm{0}_{n \times n}$. The transpose is denoted by $(\cdot)^{\intercal}$ and the Euclidean norm by $\|\cdot\|$. For a vector $\bm{v} \in \mathbb{R}^3$, $(\bm{v})_i$ denotes its $i$-th component, where $i \in \{x,y,z\}$, and $\mathbb{S}^2 \triangleq \{\bm{u} \in \mathbb{R}^3 : \|\bm{u}\| = 1\}$ denotes the unit sphere.

The skew-symmetric matrix $[\bm{a}]_{\times}$ is defined such that $[\bm{a}]_{\times}\bm{b} = \bm{a} \times \bm{b}$. Quaternions are written with the scalar part first, and the operator $\otimes$ denotes the Hamilton product of two quaternions. The exponential map $\Exp: \mathbb{R}^3 \rightarrow \mathrm{SO}(3)$, $\Exp(\bm{\phi}) = \exp([\bm{\phi}]_{\times})$, converts a rotation vector $\bm{\phi}$ into a rotation matrix, and $\Log: \mathrm{SO}(3) \rightarrow \mathbb{R}^3$ denotes its inverse. Finally, $\mathcal{N}(\bm{\mu}, \bm{\Sigma})$ denotes a Gaussian distribution with mean $\bm{\mu}$ and covariance $\bm{\Sigma}$, and $\|\bm{r}\|_{\bm{\Sigma}}^{2} \triangleq \bm{r}^{\intercal}\bm{\Sigma}^{-1}\bm{r}$ denotes the squared Mahalanobis norm.

\subsection{State Kinematics}
\label{stateKinematics}

The spacecraft attitude is parameterized by the body-to-ECEF rotation, denoted by $\bm{q} \triangleq \bm{q}_{b}^{e}$ and $\bm{R} \triangleq \bm{R}_{b}^{e} = \bm{R}(\bm{q})$. The true-state kinematics are given by~\cite{sola2017quaternion}
\begin{equation}\label{eq:TrueStateKinematics}
\begin{aligned}
    \dot{\bm{q}} &= \tfrac{1}{2}\, \bm{q} \otimes
    \begin{bmatrix} 0 \\ \bm{\omega}_{eb}^{b} \end{bmatrix}, \\
    \dot{\bm{\beta}} &= \bm{\eta}_{u},
\end{aligned}
\end{equation}
where the angular velocity $\bm{\omega}_{eb}^{b}$ is the input to the motion model, obtained from the gyroscope measurement $\tilde{\bm{\omega}}_{ib}^{b}$ and the Earth rotation rate $\bm{\omega}_{ie}^{e}$ as follows
\begin{equation}\label{eq:GyroModel}
\begin{aligned}
    \tilde{\bm{\omega}}_{ib}^{b} &= \bm{\omega}_{ib}^{b} + \bm{\beta} + \bm{\eta}_{v}, \\
    \bm{\omega}_{eb}^{b} &= \bm{\omega}_{ib}^{b} - \bm{R}^{\intercal}\bm{\omega}_{ie}^{e} \\
    &= \tilde{\bm{\omega}}_{ib}^{b} - \bm{\beta} - \bm{\eta}_{v} - \bm{R}^{\intercal}\bm{\omega}_{ie}^{e},
\end{aligned}
\end{equation}
where $\bm{\omega}_{ie}^{e} = [0 \;\; 0 \;\; \omega_E]^{\intercal}$ is the Earth rotation rate with $\omega_E = 7.292115 \times 10^{-5}$~rad/s, and $\bm{\beta} \in \mathbb{R}^3$ is the gyroscope bias modeled as a random walk. The terms $\bm{\eta}_{v}$ and $\bm{\eta}_{u}$ are zero-mean Gaussian white noises with power spectral densities $\sigma_{v}^{2}\bm{I}_3$ and $\sigma_{u}^{2}\bm{I}_3$, corresponding to the angle random walk and the rate random walk of the gyroscope, respectively.

To construct a state estimator, the true state $\bm{x} = (\bm{q}, \bm{\beta})$ is decomposed into a nominal state $\bar{\bm{x}} = (\bar{\bm{q}}, \bar{\bm{\beta}})$ and an error state $\delta\bm{x}$~\cite{sola2017quaternion}. The nominal state results from the integration of the gyroscope readings and is given by
\begin{equation}\label{eq:NominalStateKinematics}
\begin{aligned}
    \dot{\bar{\bm{q}}} &= \tfrac{1}{2}\, \bar{\bm{q}} \otimes
    \begin{bmatrix} 0 \\ \bar{\bm{\omega}}_{eb}^{b} \end{bmatrix}, \\
    \dot{\bar{\bm{\beta}}} &= \bm{0}_{3\times1},
\end{aligned}
\end{equation}
where the nominal input is the bias-compensated angular velocity
\begin{equation}\label{eq:NominalInput}
    \bar{\bm{\omega}}_{eb}^{b} = \tilde{\bm{\omega}}_{ib}^{b} - \bar{\bm{\beta}} - \bar{\bm{R}}^{\intercal}\bm{\omega}_{ie}^{e},
\end{equation}
with $\bar{\bm{R}} = \bm{R}(\bar{\bm{q}})$.

Meanwhile, the error-state vector is defined as
\begin{equation}\label{eq:ErrorState}
\delta \bm{x} \triangleq
\begin{bmatrix}
\delta \bm{\theta}^{\intercal} &
\delta \bm{\beta}^{\intercal}
\end{bmatrix}^{\intercal} \in \mathbb{R}^{6},
\end{equation}
where $\delta\bm{\theta}$ is the attitude error and $\delta\bm{\beta}$ is the gyroscope bias error. The composition $\bm{x} = \bar{\bm{x}} \oplus \delta\bm{x}$ is given by
\begin{equation}\label{eq:Composition}
\begin{aligned}
    \bm{R} &= \bar{\bm{R}}\,\Exp(\delta\bm{\theta}), \\
    \bm{\beta} &= \bar{\bm{\beta}} + \delta\bm{\beta},
\end{aligned}
\end{equation}
or equivalently $\bm{q} = \bar{\bm{q}} \otimes \delta\bm{q}$ with $\delta\bm{q} \approx [1 \;\; \tfrac{1}{2}\delta\bm{\theta}^{\intercal}]^{\intercal}$.

Substituting \eqref{eq:Composition} into \eqref{eq:TrueStateKinematics}--\eqref{eq:GyroModel}, comparing the result with \eqref{eq:NominalStateKinematics}--\eqref{eq:NominalInput}, and retaining first-order terms, the error-state kinematics are obtained as~\cite{sola2017quaternion}
\begin{equation}\label{eq:ErrorStateKinematics}
\begin{aligned}
    \dot{\delta\bm{\theta}} &= -[\bar{\bm{\omega}}_{eb}^{b}]_{\times}\,\delta\bm{\theta} - \delta\bm{\beta} - \bm{\eta}_{v} + h.o.t., \\
    \dot{\delta\bm{\beta}} &= \bm{\eta}_{u},
\end{aligned}
\end{equation}
where the higher-order terms (\textit{h.o.t.}) are ignored. Moreover, the first-order term $-[\bar{\bm{R}}^{\intercal}\bm{\omega}_{ie}^{e}]_{\times}\,\delta\bm{\theta}$, which stems from resolving the Earth rate in $\mathcal{F}_b$ with the perturbed attitude, is neglected since its magnitude scales with $\omega_E$.

In matrix form, \eqref{eq:ErrorStateKinematics} is written as
\begin{equation}\label{eq:ErrorStateDynamics}
\dot{\delta\bm{x}} = \bm{A}(t)\,\delta\bm{x} + \bm{G}\,\bm{\eta},
\end{equation}
where $\bm{\eta} \triangleq [\bm{\eta}_{v}^{\intercal} \;\; \bm{\eta}_{u}^{\intercal}]^{\intercal}$, and
\begin{equation}\label{eq:ErrorStateMatrices}
\bm{A}(t) =
\begin{bmatrix}
-[\bar{\bm{\omega}}_{eb}^{b}]_{\times} & -\bm{I}_3 \\
\bm{0}_3 & \bm{0}_3
\end{bmatrix}, \quad
\bm{G} =
\begin{bmatrix}
-\bm{I}_3 & \bm{0}_3 \\
\bm{0}_3 & \bm{I}_3
\end{bmatrix}.
\end{equation}
The nominal kinematics \eqref{eq:NominalStateKinematics} and the error-state dynamics \eqref{eq:ErrorStateDynamics} are discretized in Section~\ref{Sec:StateEstimation} to form the gyroscope factor that connects consecutive states in the factor graph.

%%

%%\subsection{Measurement Model A}

\subsection{Factor Graph Optimization}\label{Sec:FGO}

A factor graph is a bipartite graph that represents the factorization of a probability density function, where variable nodes correspond to the unknown states and factor nodes correspond to the probabilistic constraints imposed on these states by measurements and prior information~\cite{dellaert2017factor}. In this work, the variable nodes are the states $\bm{x}_k = (\bm{q}_k, \bm{\beta}_k)$ at the discrete time instants $t_k$.

Let $\mathcal{X}$ denote the set of states and $\mathcal{Z}$ the set of measurements. The maximum a posteriori (MAP) estimate of the states is
\begin{equation}\label{eq:MAP}
\hat{\mathcal{X}} = \arg\max_{\mathcal{X}} \; p(\mathcal{X} \mid \mathcal{Z}) = \arg\max_{\mathcal{X}} \; \prod_{j} f_j(\mathcal{X}_j),
\end{equation}
where each factor $f_j$ depends only on a subset of states $\mathcal{X}_j \subseteq \mathcal{X}$. Under the Gaussian noise assumption, each factor takes the form $f_j(\mathcal{X}_j) \propto \exp\bigl(-\tfrac{1}{2}\|\bm{r}_j(\mathcal{X}_j)\|_{\bm{\Sigma}_j}^{2}\bigr)$, where $\bm{r}_j$ is the residual of the factor and $\bm{\Sigma}_j$ is its covariance. Taking the negative logarithm of \eqref{eq:MAP}, the MAP estimation reduces to the nonlinear least-squares problem
\begin{equation}\label{eq:NLS}
\hat{\mathcal{X}} = \arg\min_{\mathcal{X}} \; \sum_{j} \|\bm{r}_j(\mathcal{X}_j)\|_{\bm{\Sigma}_j}^{2}.
\end{equation}

Since the attitude evolves on $\mathrm{SO}(3)$, \eqref{eq:NLS} is solved over the error states rather than over the quaternions~\cite{forster2016manifold}. Each residual is linearized around the nominal states $\bar{\mathcal{X}}$ as
\begin{equation}\label{eq:ResidualLinearization}
\bm{r}_j(\bar{\mathcal{X}}_j \oplus \delta\mathcal{X}_j) \approx \bm{r}_j(\bar{\mathcal{X}}_j) + \bm{J}_j\,\delta\mathcal{X}_j,
\end{equation}
where $\delta\mathcal{X}_j$ stacks the error states of $\mathcal{X}_j$, the operator $\oplus$ is applied to each state as in \eqref{eq:Composition}, and $\bm{J}_j = \partial\bm{r}_j / \partial\delta\mathcal{X}_j$ is evaluated at $\delta\mathcal{X}_j = \bm{0}$. Substituting \eqref{eq:ResidualLinearization} into \eqref{eq:NLS} yields the Gauss--Newton normal equations
\begin{equation}\label{eq:NormalEquations}
\bm{H}\,\delta\mathcal{X}^{*} = -\bm{g},
\end{equation}
with
\begin{equation}\label{eq:HessianGradient}
\begin{aligned}
\bm{H} &= \sum_{j} \bm{J}_j^{\intercal}\bm{\Sigma}_j^{-1}\bm{J}_j, \\
\bm{g} &= \sum_{j} \bm{J}_j^{\intercal}\bm{\Sigma}_j^{-1}\bm{r}_j(\bar{\mathcal{X}}_j),
\end{aligned}
\end{equation}
where each $\bm{J}_j$ is placed into the columns of the states it involves, and $\bm{H}$ is the information matrix. The solution $\delta\mathcal{X}^{*}$ updates the nominal states as $\bar{\mathcal{X}} \leftarrow \bar{\mathcal{X}} \oplus \delta\mathcal{X}^{*}$, after which the error states are reset to zero, and the procedure is repeated until convergence.

Since the number of states grows over time, a sliding window that retains only the $W$ most recent states, $\mathcal{X}_k = \{\bm{x}_{k-W+1}, \ldots, \bm{x}_k\}$, is used to bound the computational cost~\cite{sibley2010sliding,leutenegger2015keyframe}. When a new state is added, the oldest state $\bm{x}_m$ is marginalized rather than discarded, so that its information is retained. To this end, only the factors connected to $\bm{x}_m$ are linearized, and the resulting normal equations are partitioned as
\begin{equation}\label{eq:MargPartition}
\begin{bmatrix}
\bm{H}_{mm} & \bm{H}_{mr} \\
\bm{H}_{rm} & \bm{H}_{rr}
\end{bmatrix}
\begin{bmatrix}
\delta\bm{x}_m \\
\delta\mathcal{X}_r
\end{bmatrix}
= -
\begin{bmatrix}
\bm{g}_m \\
\bm{g}_r
\end{bmatrix},
\end{equation}
where the subscript $r$ denotes the remaining states connected to $\bm{x}_m$. Eliminating $\delta\bm{x}_m$ through the Schur complement gives
\begin{equation}\label{eq:SchurComplement}
\begin{aligned}
\bm{H}_{\mathrm{marg}} &= \bm{H}_{rr} - \bm{H}_{rm}\bm{H}_{mm}^{-1}\bm{H}_{mr}, \\
\bm{g}_{\mathrm{marg}} &= \bm{g}_{r} - \bm{H}_{rm}\bm{H}_{mm}^{-1}\bm{g}_{m},
\end{aligned}
\end{equation}
which define a marginalization prior factor on the remaining states with the residual and covariance
\begin{equation}\label{eq:MargPrior}
\begin{aligned}
\bm{r}_{\mathrm{marg}} &= \mathcal{X}_r \ominus \bar{\mathcal{X}}_r^{\mathrm{lin}} + \bm{H}_{\mathrm{marg}}^{-1}\bm{g}_{\mathrm{marg}}, \\
\bm{\Sigma}_{\mathrm{marg}} &= \bm{H}_{\mathrm{marg}}^{-1},
\end{aligned}
\end{equation}
where $\bar{\mathcal{X}}_r^{\mathrm{lin}}$ denotes the nominal states at which the marginalization is performed, and $\ominus$ is the inverse of $\oplus$, i.e., $\bm{x} \ominus \bar{\bm{x}} = [\Log(\bar{\bm{R}}^{\intercal}\bm{R})^{\intercal} \;\; (\bm{\beta} - \bar{\bm{\beta}})^{\intercal}]^{\intercal}$. The factors connected to $\bm{x}_m$ are then removed from the graph and replaced by this prior factor.

\begin{figure*}[!t]
    \centering
    \includegraphics[width=\textwidth]{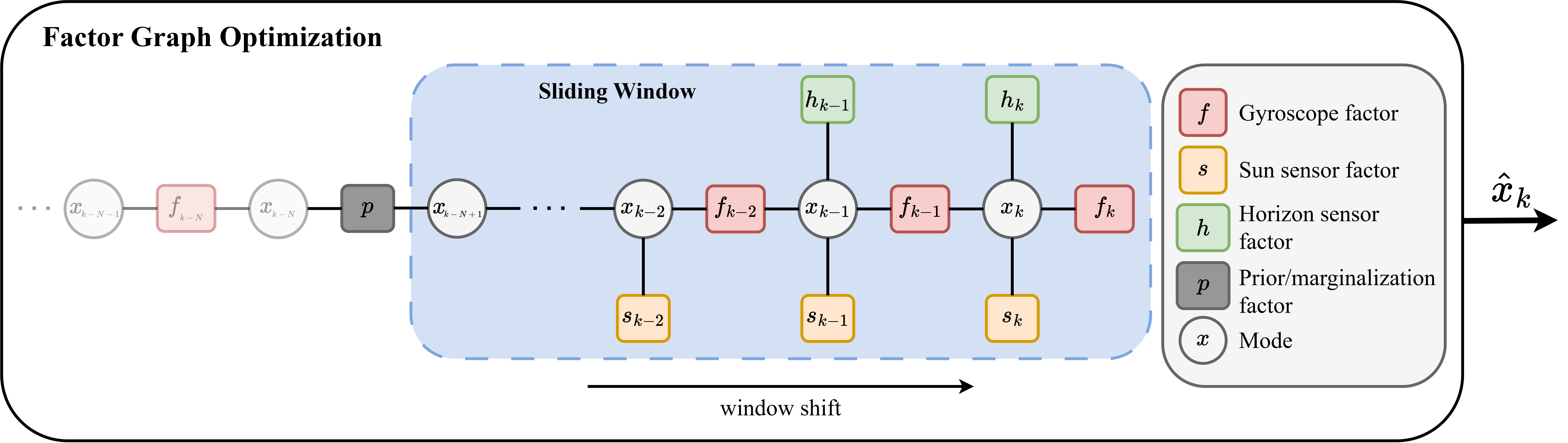}
    \caption{Sliding-window factor graph for spacecraft attitude and gyroscope-bias estimation, showing the motion factors between consecutive states and the asynchronous Sun and horizon measurement factors.}
    \label{fig:FactorGraph}
\end{figure*}

\section{Methodology}
\label{methodology}

The proposed framework consists of two stages, as illustrated in Fig.~\ref{fig:FactorGraph}. First, the horizon sensor in Section~\ref{Sec:HorizonNN} and the Sun sensor in Section~\ref{Sec:SunSensor} provide the nadir and Sun directions in $\mathcal{F}_b$, each with its own covariance. Then, the sliding-window FGO in Section~\ref{Sec:StateEstimation} fuses these measurements with the gyroscope measurements to estimate the attitude and the gyroscope bias.

\subsection{Horizon Detector Neural Net}\label{Sec:HorizonNN}

The horizon sensor provides the nadir direction in $\mathcal{F}_b$ and its covariance from a single RGB image. In this section, we first describe how the nadir direction is determined from horizon points in the image, then present the neural network that detects the horizon and predicts its localization uncertainty, and finally combine them into an uncertainty-aware nadir measurement.

\subsubsection{Horizon-Based Nadir Determination}\label{Sec:HorizonGeometry}

\begin{figure}[!b]
    \centering
    \includegraphics[width=0.85\columnwidth]{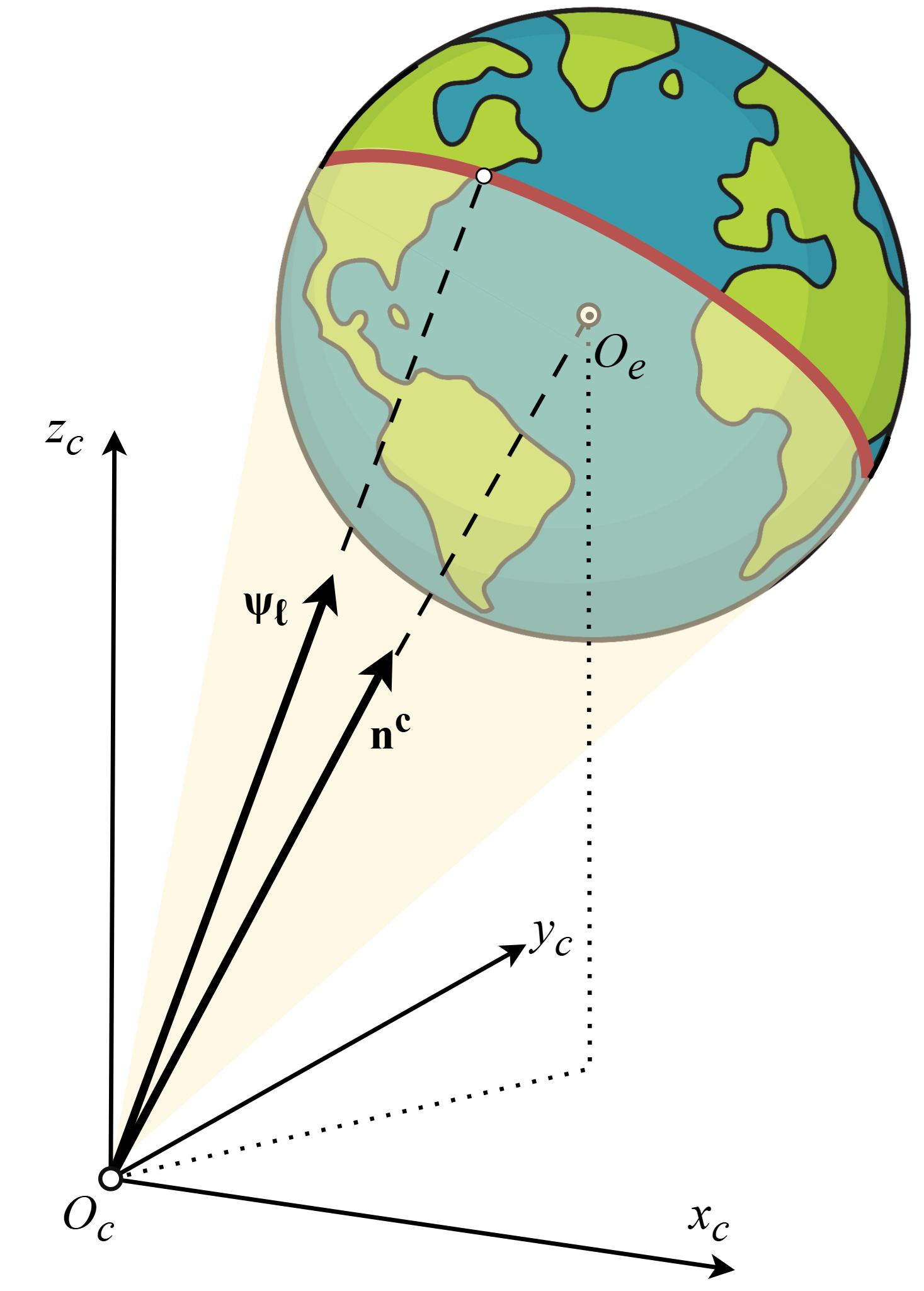}
    \caption{Geometry of horizon-based nadir determination. The camera-frame nadir direction $\bm{n}^{c}$ points from the camera origin $O_c$ toward the Earth center $O_e$, while $\bm{\psi}_{\ell}$ denotes the line-of-sight vector associated with the $\ell$-th detected horizon pixel. The collection of horizon lines of sight forms the apparent Earth-limb cone about $\bm{n}^{c}$.}
    \label{fig:earthNadir}
\end{figure}

The horizon sensor determines the nadir direction from the apparent Earth limb in a camera image~\cite{christian2021tutorial,braun2023image}. Modeling the Earth as a sphere of radius $R_E$, the lines of sight from the spacecraft that are tangent to the Earth form a right circular cone whose axis is the nadir direction, as illustrated in Fig.~\ref{fig:earthNadir}, where $O_c$ denotes the origin of $\mathcal{F}_c$ and $O_e$ the center of the Earth. The half-angle $\rho$ of this cone satisfies $\sin\rho = R_E/\|\bm{p}_{eb}^{e}\|$, where $\bm{p}_{eb}^{e}$ is the spacecraft position.

Let $\bm{u}_\ell \in \mathbb{R}^2$, $\ell = 1, \ldots, N$, denote the pixel coordinates of the detected horizon points. For a pinhole camera with focal length $f$ and principal point $\bm{c} \in \mathbb{R}^2$, both in pixels, the line of sight of the $\ell$-th horizon point resolved in $\mathcal{F}_c$ is
\begin{equation}\label{eq:LineOfSight}
\bm{\psi}_\ell = \begin{bmatrix} (\bm{u}_\ell - \bm{c})^{\intercal} & f \end{bmatrix}^{\intercal}.
\end{equation}
Since every line of sight lies on the horizon cone, the nadir direction $\bm{n}^{c}$ satisfies
\begin{equation}\label{eq:ConeConstraint}
\bm{\psi}_\ell^{\intercal}\,\bm{n}^{c} = \|\bm{\psi}_\ell\|\cos\rho, \qquad \ell = 1, \ldots, N,
\end{equation}
which is linear in $\bm{n}^{c}$~\cite{christian2016noniterative,braun2023image}. Stacking \eqref{eq:ConeConstraint} for all horizon points gives $\bm{M}\bm{n}^{c} = \bm{d}$, where $\bm{M} = [\bm{\psi}_1 \; \cdots \; \bm{\psi}_N]^{\intercal}$ and $\bm{d} = \cos\rho\,[\|\bm{\psi}_1\| \; \cdots \; \|\bm{\psi}_N\|]^{\intercal}$, with the least-squares solution
\begin{equation}\label{eq:NadirLS}
\bm{n}^{c} = (\bm{M}^{\intercal}\bm{M})^{-1}\bm{M}^{\intercal}\bm{d}.
\end{equation}
The direction of \eqref{eq:NadirLS} does not depend on $\rho$, since $\bm{d}$ scales linearly with $\cos\rho$. The measured nadir direction in $\mathcal{F}_b$ is then
\begin{equation}\label{eq:NadirBody}
\tilde{\bm{n}}^{b} = \bm{R}_{c}^{b}\,\frac{\bm{n}^{c}}{\|\bm{n}^{c}\|},
\end{equation}
where $\bm{R}_{c}^{b}$ is the known camera mounting rotation relative to body frame.

\subsubsection{Earth Segmentation Network}\label{Sec:SegNet}

The accuracy of the nadir direction depends on how precisely the horizon points are localized, which is challenging in real imagery due to clouds, atmospheric glow, and illumination changes near the terminator. To this end, the horizon is extracted from an Earth/non-Earth segmentation predicted by a neural network. We adopt a U-Net architecture~\cite{ronneberger2015u} with a ResNet-34 encoder~\cite{he2016deep} pretrained on ImageNet. For each pixel $\bm{u}$ of an RGB image, the network outputs the probability $p_{\bm{u}} \in [0,1]$ that the pixel belongs to the Earth, and the predicted mask is obtained by thresholding $p_{\bm{u}}$ at $0.5$. The network is trained with the loss
\begin{equation}\label{eq:SegLoss}
\mathcal{L}_{\mathrm{seg}} = \tfrac{1}{2}\mathcal{L}_{\mathrm{BCE}} + \tfrac{1}{2}\mathcal{L}_{\mathrm{Dice}},
\end{equation}
where the binary cross-entropy and Dice losses~\cite{milletari2016v} are
\begin{equation}\label{eq:BCE}
\begin{aligned}
\mathcal{L}_{\mathrm{BCE}} = -\frac{1}{|\mathcal{I}|}\sum_{\bm{u} \in \mathcal{I}} \bigl[\, & y_{\bm{u}}\log p_{\bm{u}} + (1 - y_{\bm{u}})\log(1 - p_{\bm{u}}) \bigr],
\end{aligned}
\end{equation}
\begin{equation}\label{eq:Dice}
\mathcal{L}_{\mathrm{Dice}} = 1 - \frac{2\sum_{\bm{u} \in \mathcal{I}} p_{\bm{u}}\,y_{\bm{u}}}{\sum_{\bm{u} \in \mathcal{I}} (p_{\bm{u}} + y_{\bm{u}})},
\end{equation}
$y_{\bm{u}} \in \{0, 1\}$ is the ground-truth label, and $\mathcal{I}$ is the set of pixels in a training batch. The cross-entropy term calibrates the per-pixel probabilities, whereas the Dice term measures the region overlap, which prevents the thin horizon boundary from being dominated by the large and easily classified interior regions.

The network is trained on a dataset of 550 RGB horizon images, consisting of 400 real images collected from publicly available sources on the internet and 150 synthetic images rendered in Unreal Engine~5~\cite{unrealengine5}. The real images cover diverse imaging conditions, such as clouds and spacecraft structures partially occluding the view. All images are manually labeled into Earth and non-Earth regions, which provide the ground-truth labels $y_{\bm{u}}$. The network is first trained on the real images and then fine-tuned on the synthetic images, mixed with real ones, to adapt it to the rendered camera images used in the simulations of Section~\ref{Sec:SimEnv}. The dataset is publicly available at \url{https://doi.org/10.5281/zenodo.22756476}.

10.5281/zenodo.22756476

\begin{figure*}[t]
    \centering
    \begin{subfigure}[b]{0.48\textwidth}
        \centering
        \includegraphics[width=\textwidth]{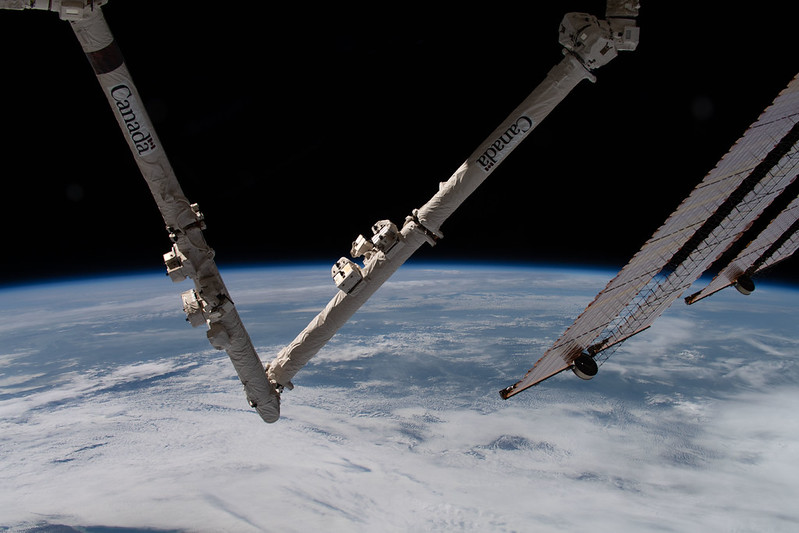}
        \caption{}
        \label{fig:left}
    \end{subfigure}%
    \hfill
    \begin{subfigure}[b]{0.48\textwidth}
        \centering
        \includegraphics[width=\textwidth]{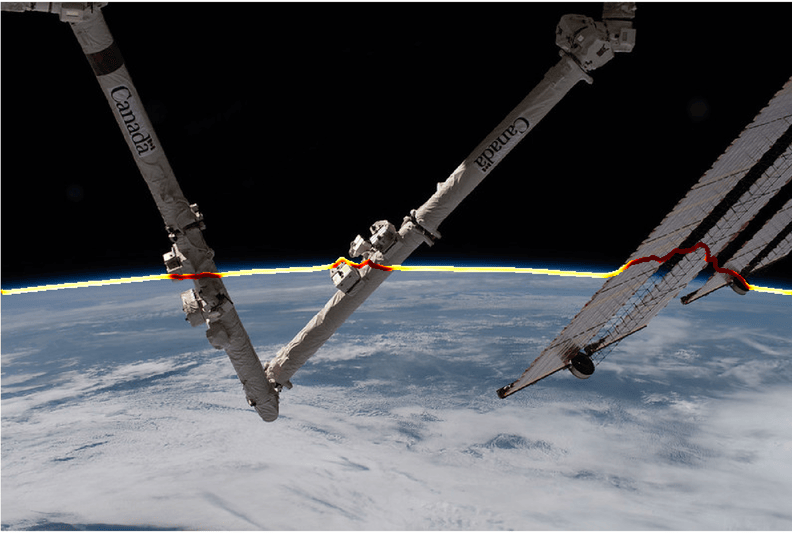}
        \caption{}
        \label{fig:right}
    \end{subfigure}
    \caption{Horizon segmentation and its predicted uncertainty on a real Earth-limb image.
(a)~Raw RGB image.
(b)~Horizon predicted by the NN, overlaid with the pixel-wise uncertainty
$\sigma$ (in pixels) in a narrow band around it. Horizon pixel colors range from white (low $\sigma$,
high confidence) to dark red (high $\sigma$, low confidence).}
    \label{fig:NNoutput}
\end{figure*}

\subsubsection{Learned Horizon Uncertainty}\label{Sec:SigmaHead}

The segmentation network provides the location of the horizon but not how accurately it is localized. Therefore, a lightweight convolutional head, fed by the final decoder features in parallel to the segmentation output, is trained to predict the per-pixel log-variance $s_{\bm{u}} = \log\sigma_{\bm{u}}^{2}$, where $\sigma_{\bm{u}}$ is the standard deviation of the horizon localization error in pixels. The regression target is obtained from signed distance maps. Let $D_{\bm{u}}^{\mathrm{pred}}$ and $D_{\bm{u}}^{\mathrm{gt}}$ denote the signed distances, in pixels, from pixel $\bm{u}$ to the boundaries of the predicted and ground-truth masks, respectively, with positive values inside the Earth region. The residual
\begin{equation}\label{eq:BoundaryResidual}
e_{\bm{u}} = D_{\bm{u}}^{\mathrm{pred}} - D_{\bm{u}}^{\mathrm{gt}}
\end{equation}
is the perpendicular offset between the predicted and true horizons, and it vanishes for a perfect prediction. The head is trained with the Gaussian negative log-likelihood loss~\cite{kendall2017uncertainties,qiu2025mac}
\begin{equation}\label{eq:SigmaLoss}
\mathcal{L}_{\sigma} = \frac{1}{|\mathcal{S}|}\sum_{\bm{u} \in \mathcal{S}} \bigl[ e_{\bm{u}}^{2}\exp(-s_{\bm{u}}) + s_{\bm{u}} \bigr],
\end{equation}
where $\mathcal{S} = \{\bm{u} \in \mathcal{I} : |D_{\bm{u}}^{\mathrm{pred}}| \leq \tau, \; |e_{\bm{u}}| \leq e_{\max}\}$ contains the pixels within a distance $\tau$ of the predicted horizon, excluding gross segmentation failures with $|e_{\bm{u}}| > e_{\max}$. The supervision is applied around the predicted horizon rather than the ground-truth horizon, since the uncertainty is queried at the predicted horizon points during inference. Moreover, the segmentation network is frozen while the head is trained. Hence, $e_{\bm{u}}$ is a fixed target, which prevents the uncertainty from being inflated on difficult pixels instead of the prediction being improved, and the segmentation accuracy remains unchanged.

The head in \eqref{eq:SigmaLoss} captures the aleatoric uncertainty of the horizon location. The epistemic uncertainty of the segmentation network can additionally be obtained by Monte Carlo (MC) dropout~\cite{gal2016dropout}. With dropout active in the decoder, $N_{\mathrm{MC}}$ stochastic forward passes produce the probability maps $p_{\bm{u}}^{(t)}$, $t = 1, \ldots, N_{\mathrm{MC}}$. Linearizing the $p_{\bm{u}} = 0.5$ level set, the resulting horizon displacement has the standard deviation $\sigma_{\bm{u}}^{\mathrm{epis}} = \mathrm{std}_t(p_{\bm{u}}^{(t)}) / \|\nabla\bar{p}_{\bm{u}}\|$, where $\mathrm{std}_t(\cdot)$ is the sample standard deviation over the passes and $\bar{p}_{\bm{u}}$ is the mean probability, and the total uncertainty becomes $\sigma_{\bm{u}}^{2} + (\sigma_{\bm{u}}^{\mathrm{epis}})^{2}$. In this work, a single forward pass ($N_{\mathrm{MC}} = 1$) is used for computational efficiency, for which $\sigma_{\bm{u}}^{\mathrm{epis}} = 0$.

Fig.~\ref{fig:NNoutput} shows the horizon and its uncertainty predicted by the network on a real Earth-limb image. The predicted uncertainty remains low along the unobstructed horizon, whereas it increases where the robotic arms and the solar array of the spacecraft occlude the limb and the predicted horizon becomes unreliable.

\subsubsection{Uncertainty-Aware Nadir Measurement}\label{Sec:NadirMeasurement}

At inference, the horizon points $\{\bm{u}_\ell\}_{\ell=1}^{N}$ are extracted from the predicted mask by tracing the boundary of the Earth region connected to the image border, splitting it at the image border, and selecting the resulting arc whose outer side is the darkest, i.e., deep space. The predicted standard deviation at each horizon point, $\sigma_\ell \triangleq \sigma_{\bm{u}_\ell}$, is then read from the uncertainty map.

The least-squares solution \eqref{eq:NadirLS} weights all horizon points equally. Instead, the predicted uncertainties are used to weight them. Let $\epsilon_\ell = \bm{\psi}_\ell^{\intercal}\bm{n}^{c} - \|\bm{\psi}_\ell\|\cos\rho$ denote the residual of \eqref{eq:ConeConstraint}. A localization error $\delta\bm{u}_\ell$ of the $\ell$-th horizon point perturbs this residual to first order as $\delta\epsilon_\ell = \bm{h}_\ell^{\intercal}\delta\bm{u}_\ell$, with
\begin{equation}\label{eq:ConstraintGradient}
\bm{h}_\ell = \frac{\partial \epsilon_\ell}{\partial \bm{u}_\ell} = \begin{bmatrix} (\bm{n}^{c})_x \\ (\bm{n}^{c})_y \end{bmatrix} - \frac{\cos\rho}{\|\bm{\psi}_\ell\|}(\bm{u}_\ell - \bm{c}).
\end{equation}
Since $\bm{h}_\ell$ is normal to the projected horizon curve, only the component of $\delta\bm{u}_\ell$ perpendicular to the horizon affects $\epsilon_\ell$, which is precisely the localization error whose standard deviation $\sigma_\ell$ is predicted by the network. Assuming independent errors among the horizon points, the covariance of the residuals is $\bm{\Sigma}_{\epsilon} = \mathrm{diag}(\sigma_1^{2}\|\bm{h}_1\|^{2}, \ldots, \sigma_N^{2}\|\bm{h}_N\|^{2})$, and the weighted least-squares solution and its covariance are
\begin{equation}\label{eq:NadirWLS}
\begin{aligned}
\bm{n}^{c} &= (\bm{M}^{\intercal}\bm{\Sigma}_{\epsilon}^{-1}\bm{M})^{-1}\bm{M}^{\intercal}\bm{\Sigma}_{\epsilon}^{-1}\bm{d}, \\
\bm{\Sigma}_{n}^{c} &= (\bm{M}^{\intercal}\bm{\Sigma}_{\epsilon}^{-1}\bm{M})^{-1}.
\end{aligned}
\end{equation}
Since $\bm{h}_\ell$ depends on $\bm{n}^{c}$, \eqref{eq:NadirWLS} is solved iteratively, starting from \eqref{eq:NadirLS}. As a result, horizon points with large predicted uncertainty contribute less to the nadir estimate, and $\bm{\Sigma}_{n}^{c}$ grows when the visible horizon arc is short or nearly straight, i.e., when $\bm{M}^{\intercal}\bm{\Sigma}_{\epsilon}^{-1}\bm{M}$ becomes ill-conditioned.

The measured nadir direction is expressed in $\mathcal{F}_b$ by \eqref{eq:NadirBody}, and its covariance by $\bm{\Sigma}_{n}^{b} = \bm{R}_{c}^{b}\,\bm{\Sigma}_{n}^{c}\,(\bm{R}_{c}^{b})^{\intercal}$. Since $\tilde{\bm{n}}^{b} \in \mathbb{S}^2$ has only two degrees of freedom, the component of $\bm{\Sigma}_{n}^{b}$ along the nadir direction represents a scale uncertainty rather than an angular error. The covariance used in the FGO is therefore obtained by projecting $\bm{\Sigma}_{n}^{b}$ onto the tangent plane of $\mathbb{S}^2$ at the nadir direction,
\begin{equation}\label{eq:NadirTangentCov}
\bm{\Sigma}_{h} = \bm{T}^{\intercal}\,\bm{\Sigma}_{n}^{b}\,\bm{T} \in \mathbb{R}^{2 \times 2},
\end{equation}
where $\bm{T} = [\bm{t}_1 \;\; \bm{t}_2] \in \mathbb{R}^{3\times 2}$, with $\bm{T}^{\intercal}\bm{T} = \bm{I}_2$, is the orthonormal tangent-plane basis that is also used to form the horizon factor residual in Section~\ref{Sec:StateEstimation}. Consequently, $\bm{\Sigma}_{h}$ varies from frame to frame with the predicted uncertainties and the visible horizon geometry.

\subsection{Sun Sensor}\label{Sec:SunSensor}

The spacecraft carries one coarse Sun sensor head on each of its six faces, so that the Sun is observed from any attitude. Each head consists of $N_d = 12$ photodiodes arranged on a polyhedral surface, whose design and angular response calibration are adopted from~\cite{cinar2025lowcost} and are not repeated here. The diode normals $\bm{n}_{k}^{b}$, $k = 1, \ldots, 6N_d$, are known in $\mathcal{F}_b$ from the mounting rotations of the heads.

Let $\bm{s}^{b} \in \mathbb{S}^2$ denote the true Sun direction in $\mathcal{F}_b$ and $\alpha_k = \arccos\bigl((\bm{n}_{k}^{b})^{\intercal}\bm{s}^{b}\bigr)$ the incidence angle of the $k$-th diode. Its normalized output follows the angular response $g(\alpha) = \max(\cos\alpha, 0)$ and is corrupted by an angular error $\nu_k \sim \mathcal{N}(0, \sigma_{\alpha}^{2})$, which lumps the mounting misalignment and the albedo-induced distortion of the incidence angle, and by a readout error $\varepsilon_k \sim \mathcal{N}(0, \sigma_{y}^{2})$ of the digitized diode current,
\begin{equation}\label{eq:DiodeModel}
\tilde{y}_k = g\bigl(\alpha_k + \nu_k\bigr) + \varepsilon_k.
\end{equation}
A diode is declared \emph{lit} and retained when its measured incidence angle $\tilde{\alpha}_k = g^{-1}(\tilde{y}_k)$ lies within the reliable half-cone $\alpha_{\max}$ of the head, which discards the grazing readings where the response is flat and the albedo contamination is strongest. 

The quantity that is linear in the Sun direction is not the diode output but its cosine, since $\cos\alpha_k = (\bm{n}_{k}^{b})^{\intercal}\bm{s}^{b}$. Inverting the response of each lit diode therefore gives a linear observation $\tilde{c}_k = \cos\tilde{\alpha}_k$, whose variance follows from propagating the two error terms of \eqref{eq:DiodeModel} to first order,
\begin{equation}\label{eq:SunObsVariance}
\sigma_{c,k}^{2} = \bigl(\sin\tilde{\alpha}_k\,\sigma_{\alpha}\bigr)^{2} + \sigma_{y}^{2},
\end{equation}
where the angular term vanishes at normal incidence, so that the readout term prevents an unbounded weight there. Stacking the lit diodes into $\bm{N} = [\bm{n}_{1}^{b} \; \cdots]^{\intercal}$ and $\tilde{\bm{c}}$, and weighting them by $\bm{W} = \mathrm{diag}(\sigma_{c,k}^{-2})$, the Sun direction and its covariance follow from a single weighted least-squares solve,
\begin{equation}\label{eq:SunWLS}
\begin{aligned}
\bm{v}^{b} &= (\bm{N}^{\intercal}\bm{W}\bm{N})^{-1}\bm{N}^{\intercal}\bm{W}\tilde{\bm{c}}, \\
\bm{\Sigma}_{v}^{b} &= (\bm{N}^{\intercal}\bm{W}\bm{N})^{-1}.
\end{aligned}
\end{equation}
All heads observe the same Sun direction, so \eqref{eq:SunWLS} pools every lit diode regardless of its face, instead of solving each head separately and averaging the resulting unit vectors as in common coarse Sun sensor implementations, which discards the cross-head geometry and the per-diode weights. A sample is accepted only when at least three diodes are lit and $\bm{N}^{\intercal}\bm{W}\bm{N}$ is well conditioned, since two lit diodes admit two mirrored solutions and nearly coplanar diode normals leave the out-of-plane component unobservable.

As in Section~\ref{Sec:NadirMeasurement}, the measured Sun direction is obtained by normalizing \eqref{eq:SunWLS},
\begin{equation}\label{eq:SunBody}
\tilde{\bm{s}}^{b} = \frac{\bm{v}^{b}}{\|\bm{v}^{b}\|},
\end{equation} and its covariance is propagated through the normalization and projected onto the tangent plane of $\mathbb{S}^2$,
\begin{equation}\label{eq:SunTangentCov}
\bm{\Sigma}_{s} = \frac{1}{\|\bm{v}^{b}\|^{2}}\,\bm{T}_{s}^{\intercal}\,\bm{P}\,\bm{\Sigma}_{v}^{b}\,\bm{P}\,\bm{T}_{s} \in \mathbb{R}^{2 \times 2},
\end{equation}
with $\bm{P} = \bm{I}_3 - \tilde{\bm{s}}^{b}(\tilde{\bm{s}}^{b})^{\intercal}$ and $\bm{T}_{s} \in \mathbb{R}^{3\times2}$ the orthonormal tangent-plane basis of the Sun factor residual in Section~\ref{Sec:SunFactor}. The removed radial component is an illumination-scale uncertainty that carries no directional information.

\subsection{State Estimation Factor Graph Optimization}\label{Sec:StateEstimation}

The attitude and the gyroscope bias are estimated using the sliding-window FGO introduced in Section~\ref{Sec:FGO}. The states are defined at node times $t_k$, which are created at a lower rate than the gyroscope sampling rate, so that a window of $W$ states spans a time interval long enough to contain horizon and Sun measurements. The set of states within the window at time $t_k$ is
\begin{equation}\label{eq:WindowStates}
\mathcal{X}_k = \{\bm{x}_{k-W+1}, \ldots, \bm{x}_k\}, \qquad \bm{x}_j = (\bm{q}_j, \bm{\beta}_j).
\end{equation}
The structure of the factor graph is shown in Fig.~\ref{fig:FactorGraph}. Consecutive states are connected by gyroscope factors, whereas horizon and Sun factors are attached to the states that correspond to the time stamps of the measurements, and the information of the marginalized states is retained by the prior factor \eqref{eq:MargPrior}. Accordingly, the states within the window are estimated by solving
\begin{equation}\label{eq:SlidingWindowCost}
\begin{aligned}
\hat{\mathcal{X}}_k = \arg\min_{\mathcal{X}_k} \Bigl\{ & \|\bm{r}_{p}\|_{\bm{\Sigma}_{p}}^{2} + \sum_{j=k-W+1}^{k-1} \|\bm{r}_{g,j}\|_{\bm{Q}_{j}}^{2} \\
& + \sum_{j \in \mathcal{K}_h} \|\bm{r}_{h,j}\|_{\bm{\Sigma}_{h,j}}^{2} + \sum_{j \in \mathcal{K}_s} \|\bm{r}_{s,j}\|_{\bm{\Sigma}_{s,j}}^{2} \Bigr\},
\end{aligned}
\end{equation}
where $\bm{r}_{g,j}$, $\bm{r}_{h,j}$, and $\bm{r}_{s,j}$ are the residuals of the gyroscope, horizon, and Sun factors with covariances $\bm{Q}_j$, $\bm{\Sigma}_{h,j}$, and $\bm{\Sigma}_{s,j}$, respectively, and $\mathcal{K}_h$ and $\mathcal{K}_s$ are the sets of state indices in the window with horizon and Sun measurements. The prior residual $\bm{r}_{p}$ is given by the initial prior $\bm{x}_0 \ominus \bm{x}_{\mathrm{init}}$ with covariance $\bm{\Sigma}_0$ until the first state is marginalized, and by the marginalization prior \eqref{eq:MargPrior} afterwards.

Whenever a new node is created, its nominal state is predicted by integrating the gyroscope measurements through \eqref{eq:NominalStateKinematics}, the oldest state is marginalized if the window is full, and the new factors are added to the graph. Then, \eqref{eq:SlidingWindowCost} is solved by the Gauss--Newton iterations in \eqref{eq:NormalEquations}, and the newest state $\hat{\bm{x}}_k$ provides the attitude estimate at $t_k$.

\subsubsection{Gyroscope Factor}\label{Sec:GyroFactor}

The gyroscope factor constrains two consecutive states $\bm{x}_k$ and $\bm{x}_{k+1}$, separated by $\Delta t = t_{k+1} - t_k$, through the attitude kinematics. Let $\tilde{\bm{\omega}}_k$ denote the mean of the gyroscope measurements over $[t_k, t_{k+1})$. Assuming a constant angular velocity over $\Delta t$, the discretization of \eqref{eq:NominalStateKinematics}--\eqref{eq:NominalInput} yields the relative rotation
\begin{equation}\label{eq:GyroDeltaR}
\Delta\tilde{\bm{R}}_k(\bm{\beta}_k) = \Exp\bigl((\tilde{\bm{\omega}}_k - \bm{\beta}_k - \bar{\bm{R}}_k^{\intercal}\bm{\omega}_{ie}^{e})\,\Delta t\bigr).
\end{equation}
The residual of the gyroscope factor is defined as
\begin{equation}\label{eq:GyroResidual}
\bm{r}_{g,k} =
\begin{bmatrix}
\Log\bigl(\bm{R}_{k+1}^{\intercal}\,\bm{R}_k\,\Delta\tilde{\bm{R}}_k(\bm{\beta}_k)\bigr) \\
\bm{\beta}_k - \bm{\beta}_{k+1}
\end{bmatrix} \in \mathbb{R}^{6},
\end{equation}
where the first component is the rotation between the attitude propagated from $\bm{x}_k$ and the attitude of $\bm{x}_{k+1}$, and the second component enforces the random-walk model of the bias. Perturbing \eqref{eq:GyroResidual} with \eqref{eq:Composition}, the Jacobians with respect to the error states are obtained to first order as
\begin{equation}\label{eq:GyroJacobian}
\begin{aligned}
\frac{\partial \bm{r}_{g,k}}{\partial \delta\bm{x}_k} &=
\begin{bmatrix}
\Delta\tilde{\bm{R}}_k^{\intercal} & -\Delta t\,\bm{I}_3 \\
\bm{0}_3 & \bm{I}_3
\end{bmatrix}, \\
\frac{\partial \bm{r}_{g,k}}{\partial \delta\bm{x}_{k+1}} &= -\bm{I}_6,
\end{aligned}
\end{equation}
where $\partial\bm{r}_{g,k}/\partial\delta\bm{x}_k$ coincides, to first order, with the state transition matrix of \eqref{eq:ErrorStateDynamics} over $\Delta t$. The covariance of the residual follows from the discretization of the process noise in \eqref{eq:ErrorStateDynamics}~\cite{sola2017quaternion}
\begin{equation}\label{eq:GyroCovariance}
\bm{Q}_k =
\begin{bmatrix}
\bm{Q}_{\theta\theta} & \bm{Q}_{\theta\beta} \\
\bm{Q}_{\theta\beta} & \bm{Q}_{\beta\beta}
\end{bmatrix},
\end{equation}
with $\bm{Q}_{\theta\theta} = (\sigma_v^{2}\Delta t + \tfrac{1}{3}\sigma_u^{2}\Delta t^{3})\,\bm{I}_3$, $\bm{Q}_{\theta\beta} = -\tfrac{1}{2}\sigma_u^{2}\Delta t^{2}\,\bm{I}_3$, and $\bm{Q}_{\beta\beta} = \sigma_u^{2}\Delta t\,\bm{I}_3$.

\subsubsection{Horizon Factor}\label{Sec:HorizonFactor}

The horizon factor relates the measured nadir direction $\tilde{\bm{n}}_k^{b}$ in \eqref{eq:NadirBody} to the attitude of the state $\bm{x}_k$. Assuming that the spacecraft position is known, the reference nadir direction in $\mathcal{F}_e$ is
\begin{equation}\label{eq:NadirReference}
\bm{n}_k^{e} = -\frac{\bm{p}_{eb}^{e}(t_k)}{\|\bm{p}_{eb}^{e}(t_k)\|},
\end{equation}
and the predicted nadir direction in $\mathcal{F}_b$ is $\bm{R}_k^{\intercal}\bm{n}_k^{e}$. Since both directions lie on $\mathbb{S}^2$, the residual is defined in the tangent plane at the nominal prediction $\bar{\bm{n}}_k^{b} = \bar{\bm{R}}_k^{\intercal}\bm{n}_k^{e}$ as
\begin{equation}\label{eq:HorizonResidual}
\bm{r}_{h,k} = \bm{T}_k^{\intercal}\bigl(\tilde{\bm{n}}_k^{b} - \bm{R}_k^{\intercal}\bm{n}_k^{e}\bigr) \in \mathbb{R}^{2},
\end{equation}
where $\bm{T}_k \in \mathbb{R}^{3\times2}$ is an orthonormal basis of the tangent plane at $\bar{\bm{n}}_k^{b}$. The projection removes the radial component which carries no directional information. Substituting \eqref{eq:Composition}, the predicted direction becomes $\bm{R}_k^{\intercal}\bm{n}_k^{e} \approx \bar{\bm{n}}_k^{b} + [\bar{\bm{n}}_k^{b}]_{\times}\,\delta\bm{\theta}_k$, which gives the Jacobian
\begin{equation}\label{eq:HorizonJacobian}
\frac{\partial \bm{r}_{h,k}}{\partial \delta\bm{x}_k} =
\begin{bmatrix}
-\bm{T}_k^{\intercal}[\bar{\bm{n}}_k^{b}]_{\times} & \bm{0}_{2\times3}
\end{bmatrix}.
\end{equation}
The covariance of the residual is the learned tangent-plane covariance $\bm{\Sigma}_{h,k}$ in \eqref{eq:NadirTangentCov} from NN. Since $\bm{\Sigma}_{h,k}$ grows with the predicted horizon uncertainties, horizon factors from degraded images receive less weight in \eqref{eq:SlidingWindowCost}, which makes the estimator robust to such measurements. Moreover, since $[\bar{\bm{n}}_k^{b}]_{\times}\bar{\bm{n}}_k^{b} = \bm{0}_{3\times1}$, the horizon factor does not constrain rotations about the nadir direction, which are instead observed through the Sun factor whenever the Sun and nadir directions are not collinear.

\subsubsection{Sun Sensor Factor}\label{Sec:SunFactor}

The Sun factor is constructed in the same way as the horizon factor. The reference Sun direction $\bm{s}_k^{e}$ is obtained from the solar ephemeris at $t_k$, and with the nominal prediction $\bar{\bm{s}}_k^{b} = \bar{\bm{R}}_k^{\intercal}\bm{s}_k^{e}$ and an orthonormal tangent-plane basis $\bm{T}_{s,k}$ at $\bar{\bm{s}}_k^{b}$, the residual and its Jacobian follow from \eqref{eq:HorizonResidual}--\eqref{eq:HorizonJacobian} with the nadir direction replaced by the Sun direction,
\begin{equation}\label{eq:SunResidual}
\bm{r}_{s,k} = \bm{T}_{s,k}^{\intercal}\bigl(\tilde{\bm{s}}_k^{b} - \bm{R}_k^{\intercal}\bm{s}_k^{e}\bigr) \in \mathbb{R}^{2},
\end{equation}
\begin{equation}\label{eq:SunJacobian}
\frac{\partial \bm{r}_{s,k}}{\partial \delta\bm{x}_k} =
\begin{bmatrix}
-\bm{T}_{s,k}^{\intercal}[\bar{\bm{s}}_k^{b}]_{\times} & \bm{0}_{2\times3}
\end{bmatrix},
\end{equation}
where $\tilde{\bm{s}}_k^{b}$ is the measured Sun direction \eqref{eq:SunBody} and the covariance of the residual is the per-sample tangent-plane covariance $\bm{\Sigma}_{s,k}$ in \eqref{eq:SunTangentCov}.

Since $[\bar{\bm{s}}_k^{b}]_{\times}\bar{\bm{s}}_k^{b} = \bm{0}_{3\times1}$, the Sun factor leaves rotations about the Sun direction unconstrained. The two direction factors are thus complementary, and their combined information is full rank in the attitude block unless the Sun and nadir directions become collinear.

\section{Results}\label{Sec:Results}

In this section, we first describe the simulation environment and then present the attitude and gyroscope bias estimation results of the proposed framework.

\subsection{Simulation Environment}\label{Sec:SimEnv}

\begin{figure}[!t]
    \centering
    \includegraphics[width=\columnwidth]{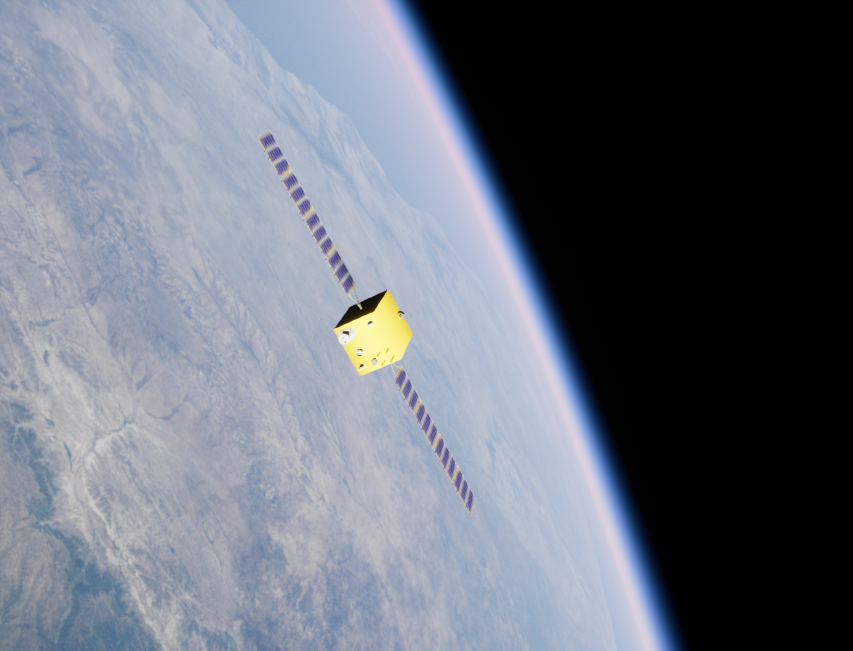}
    \caption{Earth limb and small satellite are rendered in Unreal Engine~5.}
    \label{fig:UEScene}
\end{figure}

The proposed framework is evaluated in a simulation environment developed in MATLAB/Simulink, in which the orbital motion, the attitude dynamics, and the onboard sensors of the spacecraft are modeled. The camera images are rendered in Unreal Engine~5~\cite{unrealengine5}, which is co-simulated with Simulink, as shown in Fig.~\ref{fig:UEScene}. The simulation parameters are summarized in Table~\ref{tab:SimParams}.

The spacecraft is placed in a near-circular polar orbit at an altitude of 500~km, and the simulation starts on 20 March 2021 at 09:37~UTC and lasts 500~s. The orbit is initialized such that its plane is nearly perpendicular to the Sun direction. Consequently, the Sun and nadir directions remain far from collinear, so that the horizon and Sun factors jointly constrain all three rotational degrees of freedom, as discussed in Section~\ref{Sec:SunFactor}, and the spacecraft stays in sunlight throughout the simulation.

The gyroscope measurements are generated by \eqref{eq:GyroModel} at 100~Hz, including the bias random walk, and the Sun sensor measurements are generated at 2~Hz by the model described in Section~\ref{Sec:SunSensor}. The camera captures images at 1~Hz, and its axes are aligned with $\mathcal{F}_b$, i.e., $\bm{R}_c^b = \bm{I}_3$. The nodes of the factor graph are created at 10~Hz with a window size of $W = 20$, so that each window spans 2~s. The spacecraft position required by \eqref{eq:NadirReference} is taken from the simulated orbit, and the true attitude and gyroscope bias of the simulation serve as the ground truth. The estimator is initialized with the attitude error and the gyroscope bias estimate listed in Table~\ref{tab:SimParams}.

\begin{table}[!t]
    \centering
    \caption{Simulation parameters.}
    \label{tab:SimParams}
    \footnotesize
    \begin{tabularx}{\columnwidth}{X l}
        \hline
        Parameter & Value \\
        \hline
        \multicolumn{2}{l}{\textit{Orbit}} \\
        Start epoch & 20 March 2021, 09:37 UTC \\
        Altitude, eccentricity & 500~km, 0.001 \\
        Inclination, RAAN & $90^{\circ}$, $90^{\circ}$ \\
        Argument of perigee, true anomaly & $0^{\circ}$, $0^{\circ}$ \\
        Duration & 500~s \\
        \hline
        \multicolumn{2}{l}{\textit{Gyroscope (100~Hz)}} \\
        Noise density $\sigma_{v}$ & $10^{-4}$~(rad/s)/$\sqrt{\mathrm{Hz}}$ \\
        Rate random walk $\sigma_{u}$ & $10^{-5}$~(rad/s)$\sqrt{\mathrm{Hz}}$ \\
        Initial bias & $[0.057,\, -0.057,\, 0.115]^{\intercal}$~deg/s \\
        \hline
        \multicolumn{2}{l}{\textit{Sun sensor (2~Hz)}} \\
        Angular error $\sigma_{\alpha}$ & $0.1^{\circ}$ \\
        Readout error $\sigma_{y}$ & $10^{-3}$ \\
        Reliable half-cone $\alpha_{\max}$ & $55^{\circ}$ \\
        \hline
        \multicolumn{2}{l}{\textit{Camera (1~Hz)}} \\
        Resolution & $640 \times 480$~px \\
        Horizontal field of view & $78.3^{\circ}$ \\
        Pixel error & 1~px \\
        \hline
        \multicolumn{2}{l}{\textit{Estimator}} \\
        Node rate, window size $W$ & 10~Hz, 20 \\
        Initial attitude error & $[0.6,\, -0.7,\, 0.8]^{\intercal}$~deg \\
        Initial bias estimate & $[0.020,\, -0.019,\, 0.018]^{\intercal}$~deg/s \\
        Prior standard deviation (attitude, bias) & $5^{\circ}$, 1~deg/s \\
        \hline
    \end{tabularx}
\end{table}

\subsection{Attitude Estimation Results}\label{Sec:EstimationResults}

\begin{figure}[!t]
    \centering
    \includegraphics[width=\columnwidth]{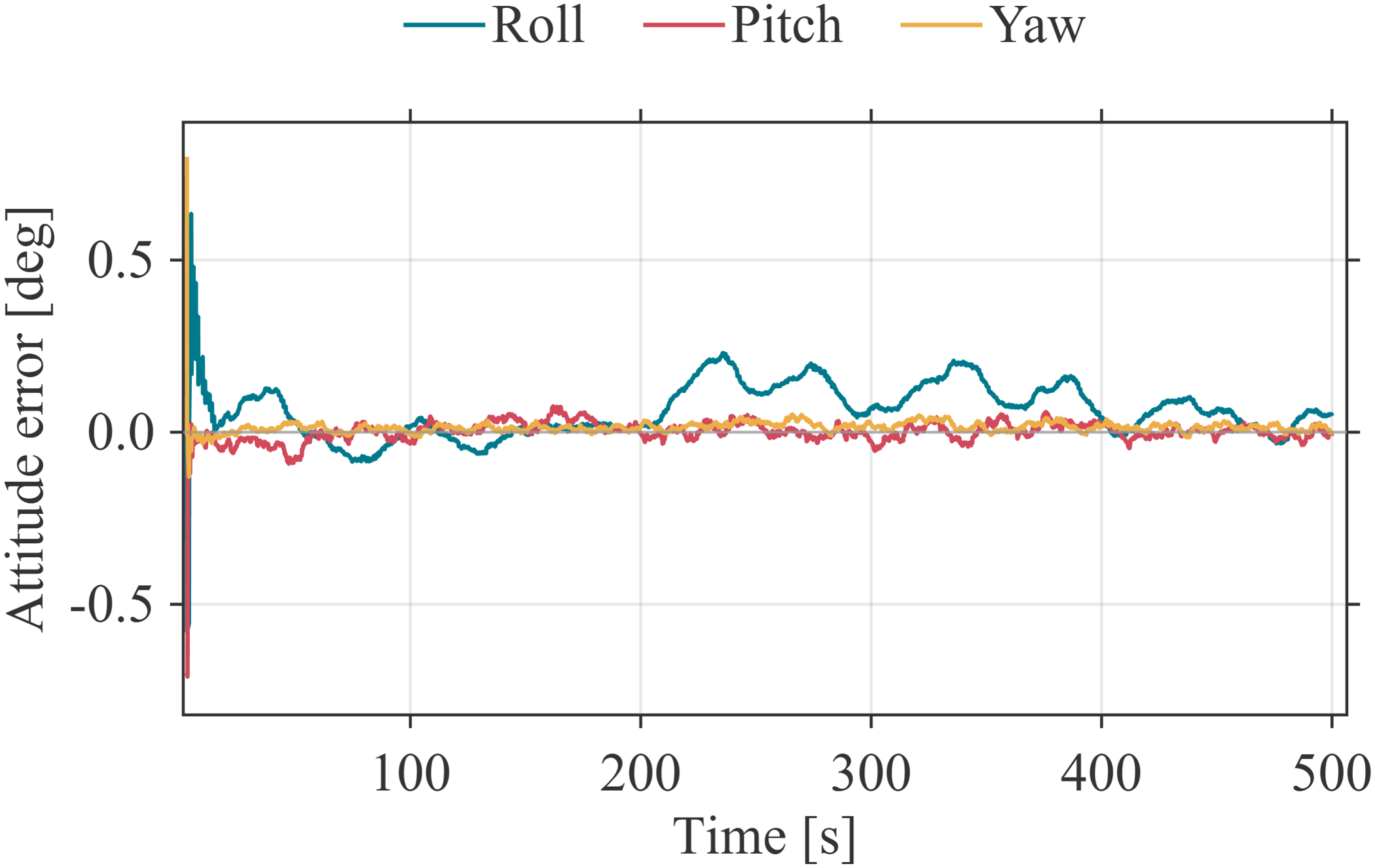}
    \caption{Attitude estimation error $\delta\bm{\theta}$ about the $x$-, $y$-, and $z$-axes of $\mathcal{F}_b$.}
    \label{fig:AttitudeError}
\end{figure}

\begin{figure}[!t]
    \centering
    \includegraphics[width=\columnwidth]{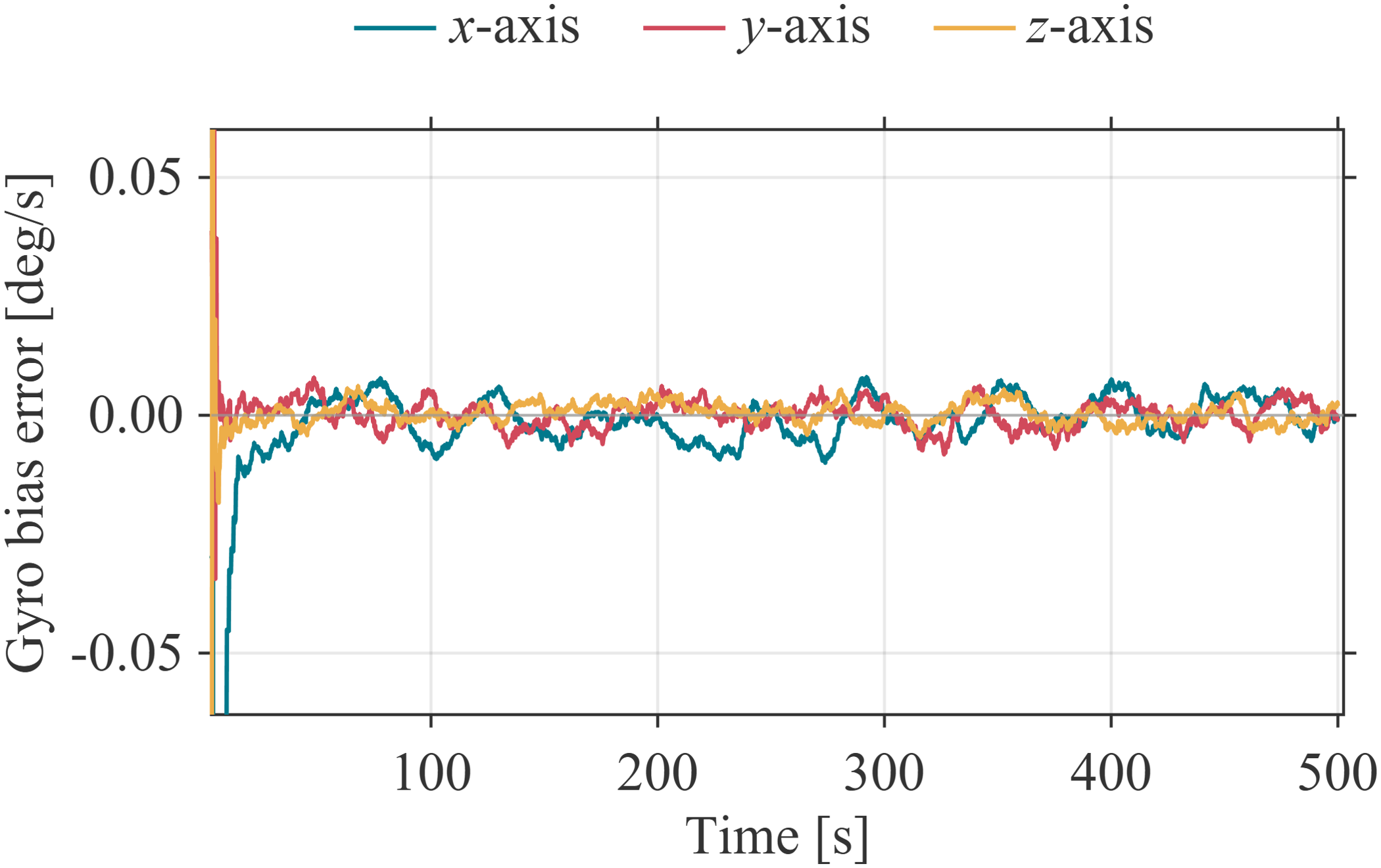}
    \caption{Gyroscope bias estimation error $\delta\bm{\beta}$ about the $x$-, $y$-, and $z$-axes of $\mathcal{F}_b$.}
    \label{fig:BiasError}
\end{figure}

Figs.~\ref{fig:AttitudeError} and~\ref{fig:BiasError} show the attitude error $\delta\bm{\theta}$ and the gyroscope bias error $\delta\bm{\beta}$ over the simulation, where the attitude error is expressed as the rotation vector between the true and estimated attitudes. Starting from the initial errors in Table~\ref{tab:SimParams}, the sliding-window FGO successfully estimates all three components of both the attitude and the gyroscope bias. This agrees with the observability discussed in Section~\ref{Sec:SunFactor}: the horizon factors constrain the rotations about the axes perpendicular to the nadir direction, the Sun factors additionally constrain the rotation about the nadir direction, and the gyroscope factors, which relate the attitudes of consecutive nodes, render the gyroscope bias observable.
% TODO: add the steady-state attitude and bias errors, e.g., RMS values after convergence.

\section{Conclusion \& Future Work}\label{Sec:Conclusion}

This paper presented a low-cost attitude determination framework for small satellites in LEO, in which a neural network detects the Earth horizon in RGB images and predicts its localization uncertainty at each pixel. These uncertainties are propagated into the covariance of the measured nadir direction, which is fused with the Sun sensor and gyroscope measurements in a sliding-window FGO, so that the attitude estimation is robust to degraded horizon measurements. Simulations with camera images rendered in Unreal Engine~5 showed that the framework successfully estimates all three components of both the attitude and the gyroscope bias.

In future work, the same learning-based approach will be applied to the Sun sensor, so that the uncertainty of each Sun measurement, e.g., due to Earth albedo, is also predicted by a neural network. Moreover, the framework will be evaluated on data recorded during a real spacecraft flight. Finally, the spherical Earth assumption will be relaxed by accounting for the Earth oblateness~\cite{deng2023large}, and the epistemic uncertainty of the network will be included through MC dropout when the onboard computational resources allow.

\bibliography{references}

\end{document}